\documentclass[preprint,amsmath,showpacs,amssymb,prb,a4paper]{revtex4}
\usepackage{graphicx,bm}

\def\unitangstrom{\,\textrm{\AA}}
\def\unitev{\,{\rm eV}}

\def\unitps{\,{\rm ps}}
\def\unitfs{\,{\rm fs}}

\def\unitpercent{\,\%}

\begin{document}

\title{Origin of coherent G band phonon spectra in single wall carbon
  nanotubes}

\author{A.~R.~T.~Nugraha$^1$, E.~H.~Hasdeo$^1$, G.~D. Sanders$^2$,
  C.~J.~Stanton$^2$, R.~Saito$^1$}

\affiliation{ $^1$Department of Physics, Tohoku University, Sendai
  980-8578, Japan\\
  $^2$Department of Physics, University of Florida, Box 118440,
  Gainesville, Florida 32611-8440, USA}

\date{\today}

\begin{abstract}
  Coherent phonons in single wall carbon nanotubes (SWNTs) are
  observed as oscillations of the differential absorption coefficient
  as a function of time by means of pump-probe spectroscopy.  For the
  radial breathing mode (RBM) of a SWNT, the coherent phonon signal is
  understood to be a result of the modulated diameter-dependent energy
  gaps due to the coherent RBM phonon oscillations.  However, this
  mechanism might not be the dominant contribution to other phonon
  modes in the SWNT.  In particular, for the G band phonons, which
  correspond to bond-stretching motions, we find that the modulation
  of the interatomic dipole matrix element gives rise to a strong
  coherent G band phonon intensity comparable to the coherent RBM
  phonon intensity.  We also further discuss the dependence of
  coherent G band and RBM phonon amplitudes on the laser excitation
  pulse width.
\end{abstract}

\date{\today}
\maketitle

\section{Introduction}
Single wall carbon nanotubes (SWNTs), characterized by the chiral
index $(n,m)$,~\cite{saito98-phys} have been important materials that
provide us with a one-dimensional model system to study the dynamics
and interactions between electrons, photons, and
phonons.~\cite{ado08-review} In particular, rapid advances in
ultrafast pump-probe spectroscopy have allowed researchers to observe
lattice oscillations of SWNTs with the same phase in real time, known
as coherent phonon
spectroscopy.~\cite{gambetta06-cp,lim06-cpexp,kato08-cpaligned,
  kim09-cpprl,makino09-cpdoping} The coherent phonon motions can be
observed as oscillations of optical properties, such as the
differential transmittance ($\Delta T/T$) or differential reflectivity
($\Delta R/ R$) as a function of delay time between pump and probe
pulses.  By performing a Fourier transform of the oscillations of
$\Delta T/T$ or $\Delta R/R$ with respect to time, we can obtain the
coherent phonon spectra as a function of phonon frequency.  Several
peaks found in the coherent phonon spectra of a SWNT correspond to
Raman active phonon modes, such as the radial breathing modes (RBMs),
D bands, G bands, and G$'$ bands.~\cite{kim13-cp} Lim {\it et al.}
showed that even the low-frequency acoustic phonon signals can be
observed in purified $(6,5)$ SWNTs by coherent phonon spectroscopy
because of its ultrafine spectral resolution.~\cite{lim14-cpfund}
Moreover, ultrafast spectroscopy techniques allow us to directly
measure phonon dynamics, including phase information, or life time of
phonons, in the time
domain.~\cite{gambetta06-cp,lim06-cpexp,kim09-cpprl}

It is known that oscillations of $\Delta T/T$ or $\Delta R /R$ as a
function of delay time $t$ between pump and probe pulses in coherent
phonon spectroscopy are directly related to the modulations of the
absorption coefficient $\alpha$ as a function of the probe energy
$E_{\rm probe}$ and $t$.~\cite{zeiger92-cp} Therefore, in order to
obtain the coherent phonon spectra theoretically, we need to calculate
the absorption coefficient $\alpha(E_{\rm probe}, t)$ for a given
coherent phonon amplitude.  In the case of RBMs, the oscillations of
$\alpha(E_{\rm probe}, t)$ have been understood as a result of energy
gap modulations, which are inversely proportional to the nanotube
diameter.~\cite{lim06-cpexp,kim09-cpprl} However, in the case of G
bands, which are assigned to longitudinal-optical (LO) and in-plane
transverse-optical (iTO) phonon modes,~\cite{ado08-review} it is known
that these modes do not significantly modify the energy gaps because
the SWNT diameters are not sensitive to the LO/iTO vibrations.  While
the coherent G band signals are experimentally observed to be on the
same order of magnitude as the RBM
signals,~\cite{lim14-cpfund,kim12-deph} our previous theoretical
calculation predicted that the modulations of absorption coefficient
due to the G band (LO) phonons are about $1000$ times smaller than
those caused by the RBM.~\cite{sanders09-cp} We expect that the reason
for the discrepancy is because we considered only the change of the
energy gap as a main contribution for the coherent G band spectra and
also the excitation pulse was too long ($50\unitfs$), whereas the G
band oscillation period is about $20\unitfs$.  This fact indicates
that a different mechanism is necessary to explain the coherent G band
intensity and that the effects of laser pulse width on the coherent
phonon intensity should be taken into account, both of which are the
main subjects of this paper.

One possible dominant contribution to the coherent G band intensity is
the modulation of electron-photon interaction.  For example, Gr\"uneis
\emph{et al.}  discussed the optical absorption of graphene from $\pi$
to $\pi^*$ bands, where the atomic dipole matrix elements for the
nearest neighbor carbon-carbon atoms, $m_{\rm opt}$, are
essential.~\cite{alex03-opt} The optical matrix elements are thus
sensitive to the change in the carbon-carbon bond length, which can be
significantly modified by the G band phonons.  In this work, in
addition to the changes in electronic structure which arise from the
coherent phonons, we now consider changes to the optical matrix
element which arise from the coherent phonon oscilllations.  We find
that modulation of $m_{\rm opt}$ is particularly relevant to the
coherent G band intensity and that the changes to the optical matrix
element for the G band are larger than for the RBM oscillations.  We
calculate the coherent G band spectra for a specified SWNT chirality
and compare them with the other coherent phonon modes in the SWNT.  By
a simple analytical model, we also study how the variation of the
laser pulse width affects the coherent phonon intensity.

This paper is organized as follows.  In Section II, we explain
coherent phonon simulation methods which include a general theory for
the generation and detection of coherent phonons in SWNTs.  In Section
III, we present the main results and discuss how the coherent G band
intensity could have a stronger signal by considering the modulation
of optical interaction and shorter pulse width.  And finally, we give
a conclusion in Section IV.

\section{Simulation methods}
\label{sec:theory}

To calculate coherent phonon spectra, we follow the methods described
in our earlier papers,~\cite{sanders09-cp,nugraha11-cp} except that we
will also now treat the effects of the coherent phonon modulations of
the electron-photon interaction which we previously neglected for
simplicity.  We define a coherent phonon mode with wavevector $q = 0$
($\Gamma$ point phonon) whose amplitude satisfies a driven oscillator
equation
\begin{equation}
\label{Coherent phonon driven oscillator equation}
\frac{\partial^2 Q_{m}(t)}{\partial t^2} +
\omega^2_m Q_{m}(t) = S_m(t) ,
\end{equation}
where $m$ and $\omega_m$ denote the phonon mode (e.g. RBM, oTO, LO,
iTO) and its frequency, respectively.  Equation~\eqref{Coherent phonon
  driven oscillator equation} is solved subject to the initial
conditions $Q_m(0) = 0$ and $\dot{Q}_m(0) = 0$.  The driving function
$S_m(t)$ in the right hand side of Eq.~\eqref{Coherent phonon driven
  oscillator equation} is given by
\begin{equation}
S_m(t) = -\frac{2\omega_m}{\hbar} \sum_{nk}
{\cal M}^m_{n}(k) \left( f_{n}(k,t) - f^0_{n}(k) \right).
\label{eq:cpdrive}
\end{equation}
where $f_{n}(k,t)$ is the time-dependent electron distribution
function and $f^0_{n}(k)$ is the initial equilibrium electron
distribution function.  Here $n$ labels an electronic state, while $k$
gives the electron wavevector.  The electronic states of a SWNT are
calculated within the extended tight-binding (ETB)
approximation.~\cite{georgii04-apl} The electron-phonon matrix element
${\cal M}^m_{n}(k)$ in Eq.~\ref{eq:cpdrive} is a shorthand for ${\cal
  M}^{m,0}_{nk;nk}$, where ${\cal M}^{m,q}_{n'k';nk}$ is the
deformation potential electron-phonon matrix element in the ETB model
with phonon wavevector $q = k - k'$ and with a transition from the
state $n$ to $n'$.~\cite{jiang05-elph}. 

From Eq.~\eqref{eq:cpdrive}, we see that the driving function $S_m(t)$
depends on the photoexcited electron distribution functions, which can
be calculated generally by taking photogeneration and relaxation
effects into account.  In coherent phonon spectroscopy, an ultrafast
laser pulse generates electron-hole pairs on a time scale short in
comparison with the coherent phonon period.  The observed coherent
phonon intensity is then proportional to the power spectrum of the
oscillations of optical properties.~\cite{zeiger92-cp} Within the
scope of this work, we ignore relaxation effects of the photoexcited
carriers and consider only the rapidly varying photogeneration term
which can be calculated directly from the Fermi's golden rule.
Neglecting carrier relaxation has a negligible effect on the computed
coherent phonon signal since the relaxation time is much greater than
the laser pulse duration and the coherent phonon
period.~\cite{sanders09-cp} Using the Fermi's golden rule, we obtain
the photogeneration rate for the distribution
functions,~\cite{chuang95}
\begin{eqnarray}\label{eq:rate}
  \nonumber &&
  \frac{\partial f_{n}(k)}{\partial t} =
  \frac{8 \pi^2 e^2 \ u(t)}{\hbar \ n_g^2 \ (E_{\rm pump})^2}
  \left(\frac{\hbar^2}{m_0} \right) \sum_{n'}
  \left| P_{n n'}(k,t) \right|^2
  \\ &&
  \times \Big( f_{n'}(k,t) - f_{n}(k,t) \Big)
  \ \delta \Big( E_{n n'}(k,t) - E_{\rm pump} \Big) ,
\end{eqnarray}
where $E_{n n'}(k,t) = \arrowvert E_{n}(k,t) - E_{n'}(k,t) \arrowvert$
are the $k$ dependent transition energies at time $t$ of a coherent
phonon oscillation, $E_{\rm pump}$ is the pump laser energy, $u(t)$ is
the time-dependent energy density of the pump pulse, $e$ is the
electron charge, $m_0$ is the free electron mass, and $n_g$ is the
refractive index of the surrounding medium.  The pump energy density
$u(t)$ is related with the pump fluence $F$ by a relation $F = (c/n_g)
\int u(t) dt$ and $u(t)$ is also assumed to be a Gaussian.  Thus
\begin{align}
\label{eq:pulse}
u(t) = A_p e^{-4 t^2 \ln 2 /2\tau_p^2} ,
\end{align}
where $A_p = (2 n_g F \sqrt{\ln 2 / \pi})/ (c\tau_p)$, with $c$ is the
speed of light.  In Eq.~\eqref{eq:pulse}, $\tau_p$ is defined as the
pump duration or laser pulse width.  Unless otherwise mentioned, we
use parameters $\tau_p = 10\unitfs$, $F=10^{-5}~{\rm J cm}^{-2}$, and
$n_g = 1$.  To also account for spectral broadening of the laser
pulses, we replace the delta function in Eq.~(\ref{eq:rate}) with a
Lorentzian lineshape
\begin{equation}\label{eq:delta}
  \delta(E_{nn'} - E_{\rm pump}) \rightarrow
  \frac{\Gamma_p /(2\pi)} {{(E_{nn'} - E_{\rm pump})^2+(\Gamma_p/2)^2}} ,
\end{equation}
where $\Gamma_p = 0.15\unitev$ is the spectral linewidth (FWHM) of the
pump pulse.~\cite{sanders09-cp}

By considering light polarized parallel to the tube axis ($z$ axis)
that contributed to the optical absorption, we can write the optical
matrix element $P_{nn'}$ in Eq.~\eqref{eq:rate} within the dipole
approximation as~\cite{alex03-opt}
\begin{equation}\label{eq:optmat}
  P_{n n'}(k)= \frac{ \hbar}{\sqrt{2 m_0}} \sum_{i,jN}
  C_{i}^* (n', k) C_j(n,k) e^{i\phi_N(k)} m_{\rm opt}
  (i,jN),
\end{equation}
where $C_i (n,k)$ and $\phi_N(k)$ respectively denote the expansion
coefficient and phase factor from the $N$th two-atom unit cell of the
symmetry-adapted ETB wave functions.~\cite{popov04-opt} The atomic
dipole matrix element is given by
\begin{equation}\label{eq:mopt}
  m_{\rm opt} = \int d\textbf r \varphi_{i0}^* (\textbf{r} -
  \textbf{R}_{i0}) \frac{\partial}{\partial z} \varphi_{jN}
  (\textbf{r} - \textbf{R}_{jN}),
\end{equation}
where $\phi_{i,N}$ is the $2p_z$ orbital of the $i$th atom in
the $N$th unit cell.

We should note that Eqs.~\eqref{eq:optmat} and \eqref{eq:mopt} still
do not have an explicit time dependence.  The time-dependence of the
optical matrix element comes from the coherent phonon amplitude
$Q_m(t)$ which allows the atomic matrix element $m_{\rm opt}$ to also
vary as a function of time as the positions of the carbon atom change.
Gr\"uneis \emph{et.~al} calculated the integral in Eq.~\eqref{eq:mopt}
for planar graphene analytically by expanding the orbital
wavefunctions in terms of Gaussians and it was found that $m_{\rm
  opt}$ explicitly depends on the bond length between two carbon atoms
$a_{\rm CC}$.~\cite{alex03-opt} If the bond length $a_{\rm CC}$ is
altered by coherent phonon oscillations, the atomic matrix element
$m_{\rm opt}$ is directly affected, as is the dipole optical matrix
element $P_{nn'}$.  This is because the deformation of the bond
lengths alters the transfer integral and overlap matrix elements in
the ETB model.

Based on above argument, the time-dependence of $E_{nn'}(k,t)$ and
$P_{nn'}(k,t)$ can be obtained from the time-dependent lattice
displacements due to the change in $a_{\rm CC}$ by the coherent phonon
oscillations, especially for the G band, which is the in-plane C-C
bond-stretching mode.  From the coherent phonon amplitudes, the
time-dependent macroscopic displacements of each carbon atom in an
SWNT are given by
\begin{equation} \label{Coherent phonon displacement}
  \textbf{U}_{r,N}(t) = \frac{\hbar}{\sqrt{2 M}} \sum_{m}
  \frac{\hat{\textbf{e}}_{r\vec{J}}^m}{\sqrt{\hbar\omega_m}}
  \ Q_{m}(t)
\end{equation}
where $\hat{\textbf{e}}_{r\vec{J}}^m \equiv
\hat{\textbf{e}}_{sj}^m(q=0)$, $\hbar\omega_m \equiv
\hbar\omega_m(q=0)$, and $M$ is the mass of a carbon atom.  The bond
length $a_{\rm CC}$ at each time $t$ of a coherent phonon oscillation
can then be calculated from the macroscopic carbon atom displacements.
Therefore, the time-dependent optical matrix element can be evaluated
by
\begin{equation}
  P_{nn'} (k,t) = P_{nn'}(k,0) + \Delta P_{nn'} (k,t),
\end{equation}
where $\Delta P_{nn'}(k,t)$ is directly proportional to the
time-dependent $m_{\rm opt}$ and we take an average of $P_{nn'}$ over
three nearest neighbor atoms.

In coherent phonon spectroscopy, a laser probe pulse is used to
measure the time-varying absorption coefficient of the SWNT. The
time-dependent absorption coefficient $\alpha (t)$ at a probe energy
$E_{\rm probe}$ is given by the Fermi's golden rule
\begin{align} \label{Absorption coefficient}
  \alpha(E_{\rm probe},t)
  \propto \sum_{n n'} &\int dk \ \arrowvert P_{n
    n'}(k,t) \arrowvert^2 \Big( f_{n}(k,t) - f_{n'}(k,t) \Big)
  \nonumber\\
 &\times \delta \Big(E_{nn'}(k,t) - E_{\rm probe} \Big) ,
\end{align}
We replace the delta function in Eq.~(\ref{Absorption coefficient})
with a broadened Lorentzian spectral lineshape with a FWHM of
$\gamma=0.15~\unitev$,~\cite{sanders09-cp} similar to that in
Eq.~\eqref{eq:delta}.  Excitation of coherent phonons by the laser
pump modulates the optical properties of the SWNTs, which gives rise
to a transient differential transmission signal, or the modulations of
absorption coefficient.  The time-resolved differential gain measured
by the probe is then given by
\begin{equation}
\label{eq:gain}
  \Delta \alpha (E_{\rm probe}, t) =
  -[\alpha (\hbar \omega, t) - \alpha (\hbar \omega, t\rightarrow -\infty)]
\end{equation}
We take the theoretical coherent phonon signal (or intensity, $I$) to
be proportional to the Fourier power spectrum of such absorption
modulations at a given energy $E_{\rm probe}$,
\begin{equation}
I(\omega) = \int e^{-i\omega t} \left|\Delta \alpha (E_{\rm probe},t) \right|^2 dt,
\end{equation}
where $\omega$ represents the phonon frequency that contributes to the
coherent phonon spectra.

\section{Results and discussion}
\subsection{Modulation of optical interaction}

\begin{figure*}[t!]
  \centering\includegraphics[clip,width=17cm]{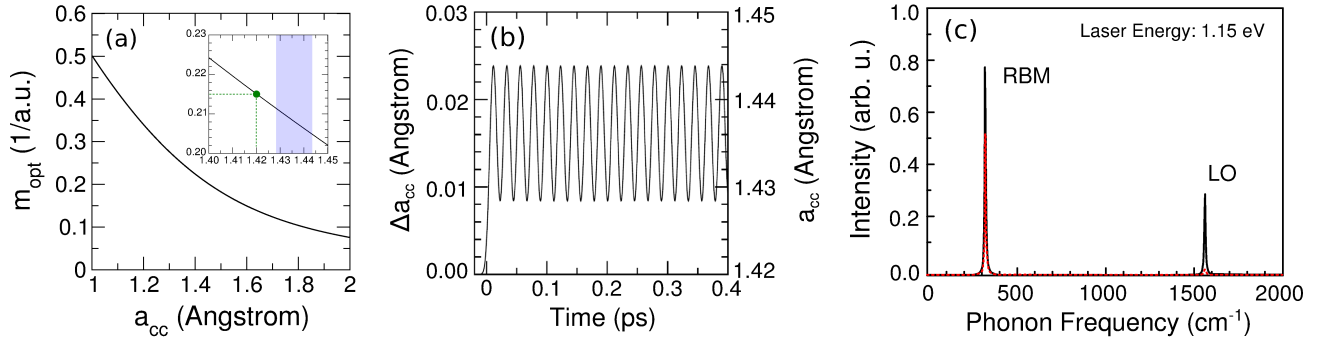}
  \caption{\label{fig1} (a) Atomic matrix element as a function of
    carbon-carbon bond length in SWNTs and (b) the change of bond
    length as a function of time due to a coherent LO phonon
    oscillation of in a $(6,5)$ SWNT at a laser (both pump and probe)
    energy of $1.15\unitev$.  Inset in (a) shows an enlarged region
    between $1.40\unitangstrom$ and $1.45\unitangstrom$.  A dot in the
    inset corresponds to the bond length without coherent phonon
    oscillations, whereas the shaded region corresponds to the area in
    which the bond length oscillates as shown in (b). (c) Coherent
    phonon intensity as a function of phonon frequency, showing the
    RBM and LO peaks.  Solid (dashed) line is the calculated result
    with (without) considering the modulation of optical matrix
    element. }
\end{figure*}

First we discuss the effects of coherent phonon oscillations on the
optical interaction.  The changes in $a_{\rm CC}$ modulate the atomic
matrix element $m_{\rm opt}$ because of the direct correspondence
between these two quantities at time $t$.  Fig.~\ref{fig1}(a) shows
the calculated $m_{\rm opt}$ as a function of $a_{\rm CC}$ based on
the formula given by Gr\"uneis \emph{et
  al.}~\cite{alex03-opt,alex04-thesis}.  It indicates that the
strength of optical interaction monotonically decreases as a function
of $a_{\rm CC}$.  In the inset of Fig.~\ref{fig1}(a), we show the
atomic matrix element within an enlarged region around
$1.40\unitangstrom$ and $1.45\unitangstrom$.  The shaded region
corresponds to the possible values of $a_{\rm CC}$ affected by the
coherent LO phonon oscillation given in Fig.~\ref{fig1}(b).  From this
figure, we can say that the modulations of optical interaction is
about $0.02$ [a.u]$^{-1}$ for the change of vibration amplitude of
about $0.02\unitangstrom$.  These modulations of optical interaction
is thus approximately $10\unitpercent$ of $m_{\rm opt} =
0.25$[a.u.]$^{-1}$, which is not negligible for calculating the
absorption coefficient of a SWNT.  The coherent phonon intensity is
proportional to $|\Delta \alpha|^2 \propto |\Delta m_{\rm opt}|^4$,
which is the leading order of the spectra.  In the previous study,
however, this fact was not taken into account and the optical matrix
element was considered constant as a function of
time.~\cite{sanders09-cp}

Next, from the time-dependent optical matrix elements, we proceed to
the calculation of coherent phonon spectra by taking the Fourier
transform of Eq.~\eqref{eq:gain}.  The calculation is performed by
allowing the probe energy in Eq.~\eqref{Absorption coefficient} to be
varied independently while keeping the pump energy in
Eq.~\eqref{eq:rate} constant.  We take a $(6,5)$ SWNT chirality as a
sample for this calculation.  This SWNT has the first and second
optical transition energies (band gaps) of $1.27\unitev$ and
$2.42\unitev$, denoted by $E_{11}$ and $E_{22}$,
respectively.~\cite{nugraha10-env} In this calculation we neglect the
exciton effects for simplicity.  Basically, the exciton-photon matrix
elements are about 100 times larger than the electron-photon matrix
elements,~\cite{jiang07-exphop} but such enhancement factors are
common for all the phonon modes.  Therefore, the exciton effects will
not modify the relative intensity between the phonon modes.  In
Fig.~1(c), we show an example of the calculation for intensity as a
function of phonon frequency by including or excluding the modulation
of optical interaction.  The coherent phonon spectra shows both the
RBM and LO peaks for a particular laser excitation energy of
$1.15\unitev$.  It can be seen that the LO intensity is enhanced
significantly when taking the modulation of optical interaction into
account, while the RBM intensity is just enhanced slightly.

\begin{figure}[t]
  \centering
  \includegraphics[clip,width=8cm]{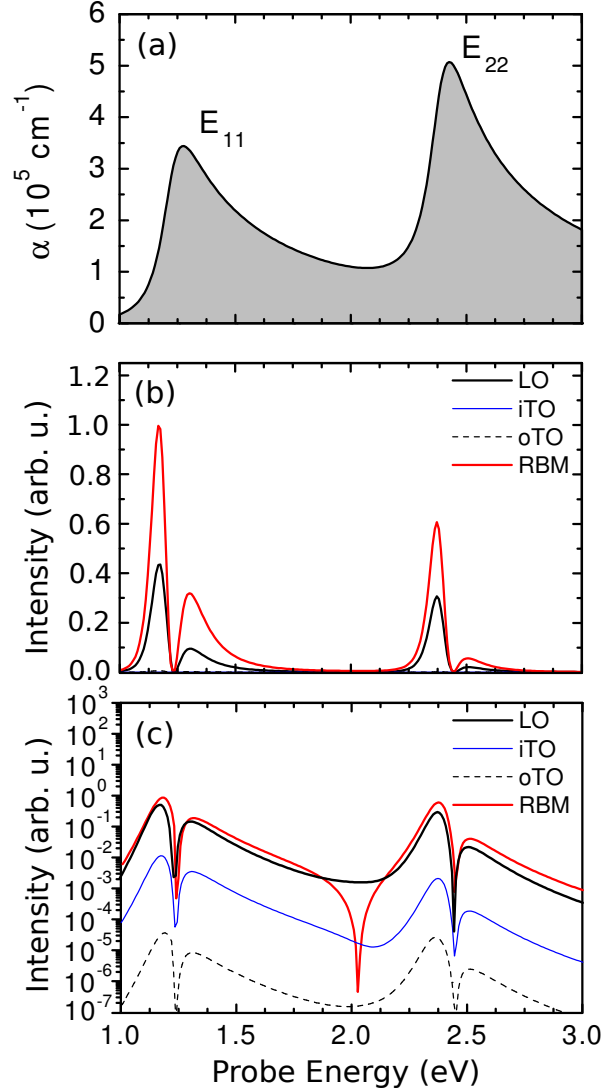}
  \caption{\label{fig2} (Color online) From top to bottom shows (a)
    absorption coefficient, (b) linearly scaled and (c)
    logarithmically scaled coherent phonon intensities as a function
    of probe energy for a (6,5) SWNT.  The intensity is normalized to
    the maximum intensity of the RBM.}
\end{figure}

To further understand the laser energy dependence of the spectra, we
calculate the coherent phonon intensity for a given phonon frequency
$\omega_m$ by considering different laser probe energy from
$1.0$-$3.0\unitev$ with an interval of $0.1\unitev$.  In
Fig.~\ref{fig2}, we show absorption coefficients and coherent phonon
spectra of the $(6,5)$ SWNT as a function of probe energy.  For the
coherent phonon spectra, we give the spectra both in the linear scale
and logarithmic scale as shown in Figs.~\ref{fig2}(b) and (c),
respectively.  The spectra are accompanied with the plot of absorption
coefficient in Fig.~\ref{fig2}(a) as a reference for showing the
positions of the optical transition energy peaks.  In
Figs.~\ref{fig2}(b) and (c), we compare the coherent G band phonon
spectra (LO and iTO modes) with RBMs and also with oTO (out-of-plane
TO) mode for the $(6,5)$ tube.  In Fig.~\ref{fig2}(b), we can see that
the coherent RBM intensity and LO intensity are on the same order,
with the RBM intensity being slighly larger than that of the LO
intensity by a ratio of about $2.5$ and $2.1$ at $E_{11}$ and
$E_{22}$, respectively.  These results indicate that modulations of
optical matrix elements become important in enhancing the coherent G
band intensity.  It should be noted that the coherent iTO intensity is
hundred times smaller than the LO intensity.  Therefore, the coherent
G band phonon spectra are mainly dominated by the LO phonon modes.

It is also interesting to see in Fig.~\ref{fig2}(c) that there is a
dip at $2\unitev$ for the RBM phonons, which might be related with the
zero value of the coherent RBM phonon amplitude.~\cite{nugraha11-cp}
The dip of RBM coherent phonon spectra could give information of
photon energy that would correspond to the transition from expansion
to contraction (or vice versa) of the SWNT
diameter.~\cite{nugraha11-cp} Moreover, we obtain two peaks at each
transition energy for all phonon modes, which are consistent with some
earlier works that reported the excitation excitation energy
dependence of coherent phonon intensity always shows a derivative-like
behavior of the absorption coefficient.~\cite{lim06-cpexp,kim12-deph} The
double-peak feature at each transition energy can be symmetric or
asymmetric depending on whether or not the excitonic effects is taken
into account.~\cite{nugraha13-cpexc}  In this work, the double-peak
lineshapes are asymmetric because we neglect the exciton effects for
simplicty.  It is then worth comparing the ratio of the coherent RBM
and LO intensity obtained in this study with that in the experiment.
For example, a pump-probe measurement by Lim~{\it et al.}  gave the
RBM intensity of about eight times larger than the LO
intensity.~\cite{lim14-cpfund} Although this discrepancy is not
significant in our present discussion, we expect that it might come
from the additional effect of the selection of laser pulse width
$\tau_p$ as we will discuss below.

\subsection{Effects of laser pulse width}
To discuss the effects of laser pulse width ($\tau_p$) on the coherent
phonon intensity, we can analytically model the driving function
$S_m(t)$ of Eq.~\eqref{eq:cpdrive} by using the laser pulse in the
form of Eq.~\eqref{eq:pulse} and then solve for $Q_m(t)$.  By
understanding the $\tau_p$ dependence of $Q_m$, we can qualitatively
explain the trend of the coherent phonon intensity when $\tau_p$ is
varied.  As we can see from Eq.~\eqref{eq:cpdrive}, $S_m(t)$ is
proportional to the carrier density, $f_n(k,t)$, which can be obtained
by integrating Eq.~\eqref{eq:rate} with respect to time.  For
simplicity, we can write $S_m(t)$ to be directly proportional to the
integration of $u(t)$,
\begin{align}
  S_m(t) &\propto \int_{-\infty}^t A_p e^{-4 t'^2 \ln 2 /2\tau_p^2}
  dt'
  \notag\\
  &\propto \frac{n_g F}{2 c} \sqrt{\frac{\pi}{\ln 2}} \left[1 + {\rm
      erf}\left(\frac{2t \ln 2}{\tau_p}\right) \right],
\label{eq:force}
\end{align}
where ${\rm erf}(x) = (2/\sqrt{\pi}) \int_0^x e^{-x'} dx'$ is the
error function.  To obtain the full equality between the left-hand and
right-hand sides of Eq.~\eqref{eq:force}, we can put an additional
term of electron-phonon matrix element as also indicated in
Eq.~\eqref{eq:cpdrive}.  This additional term along with the prefactor
in the right-hand side of Eq.~\eqref{eq:force} form a constant $A_m$,
which will change only when we have different phonon modes $m$.  We
can finally write the driving function as
\begin{equation}
  S_m(t) = \frac{A_m}{2} \left[1 + {\rm
      erf}\left(\frac{2t \ln 2}{\tau_p}\right) \right],
\label{eq:finalS}
\end{equation}
and the corresponding solution for $Q_m(t)$ with an initial condition
of $Q_m(0) = 0$ and $\dot{Q}_m(0) = 0$ is
\begin{equation}
  Q_m (t) = \frac{A_m}{\omega_m^2}
  \left[1 + e^{-\omega_m^2 \tau_p^2 / 16 \ln 2} \cos(\omega_m t)
  \right] .
\label{eq:finalQ}
\end{equation}

\begin{figure}[t!]
  \centering
  \includegraphics[clip,width=8cm]{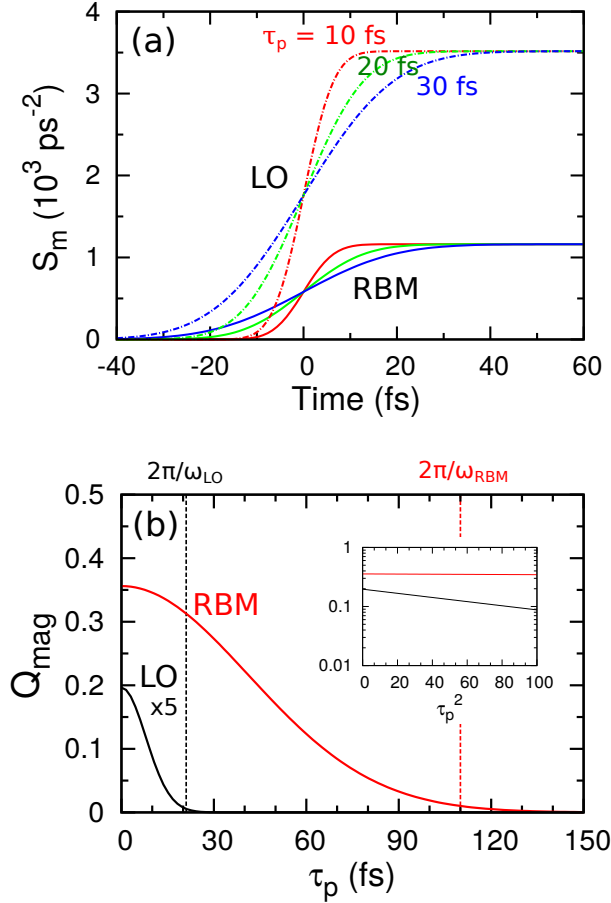}
  \caption{\label{fig3} (Color online) (a) Coherent phonon driving for
    the RBM and LO phonon modes of the $(6,5)$ SWNT under $E_{11}$
    excitation, each calculated with three different values of
    $\tau_p$: $10\unitfs$, $20\unitfs$, and $30\unitfs$.  (b) Coherent
    phonon amplitude for the RBM and LO phonon modes obtained
    analytically as a function of pulse width.  Inset shows the
    logarithmic plot of the same amplitude as a function of squared
    pulse width.}
\end{figure}

Having the solution of $Q_m(t)$, we can now discuss its dependence on
$\tau_p$.  First, the value of $A_m$ in Eqs.~\eqref{eq:finalS}
and~\eqref{eq:finalQ} can be obtained by fitting to the maximum value
of the force $S_m(t)$ simulated from the full microscopic treatment in
Eq.~\eqref{eq:cpdrive}.  In Fig.~\ref{fig3}(a), we show the simulated
$S_m(t)$ for the RBM and LO mode of $(6,5)$ SWNT under $E_{11}$
excitation with three different values of $\tau_p$.  We see that
$S_m(t)$ for all cases show a step-like behavior with a width of
$\tau_p$, and thus consistent with Eq.~\eqref{eq:finalS}.  The maximum
values of $S_m(t)$ only differ between different phonon modes.  The
fitted values of $A_m$ for the RBM and LO mode are
$1161.7\unitps^{-2}$ and $3516.5\unitps^{-2}$, respectively.  In
Fig.~\ref{fig3}(b), we show the magnitude of $Q_m$, denoted as $Q_{\rm
  mag}$, as a function of pulse width $\tau_p$ for the RBM and LO
phonon modes.  The definition for $Q_{\rm mag}$ is
\begin{equation}
  Q_{\rm mag} = \frac{A_m}{\omega_m^2}e^{-\omega_m^2 \tau_p^2 / 16 \ln 2},
\end{equation}
which represents the difference between the maximum and minimum values
of the coherent phonon oscillation amplitudes.  We also have
$2\pi/\omega_{\rm RBM} = 110\unitfs$ and $2\pi/\omega_{\rm LO} =
21\unitfs$ for the RBM and LO oscillation periods of the $(6,5)$ SWNT,
respectively.  We can see from Fig.~\ref{fig3}(b) that as the pulse
width increases, the coherent phonon amplitude quickly decays
following the Gaussian shape of the spectrum of the laser
pulse.~\cite{note14-cp} However, the rate of the amplitude decay
depends on the phonon mode oscillation frequency or period, as clearly
shown in the inset of Fig.~\ref{fig3}(b).  If the pulse width is much
smaller than the phonon oscillation period, the amplitude will be
enhanced.  In this case, the LO phonon mode is enhanced more quickly
than the RBM mode after the pulse width becomes shorter than the LO
oscillation period.  Therefore, as we have used $\tau_p = 10\unitfs$
in the simulation discussed earlier, the coherent LO intensity rapidly
increase while at the same time the coherent RBM intensity increases
more slowly.  This could be the reason why we have a slightly
different ratio of the RBM to the LO intensity since the coherent LO
phonon amplitude is much more sensitive to the variation of the laser
pulse width within sub-10 fs region compared to the coherent RBM
phonon amplitude.

\section{Conclusion}
\label{sec:conclude}
We have presented the mechanism by which a strong coherent G band
signal could be generated in ultrafast pump-probe spectroscopy.
Instead of the energy gap modulation mechanism which is dominant in
the RBM case, we suggest that the modulations of electron-photon
interaction as a function of time should be relevant to the coherent G
band intensity.  We also find an analytical formula that describes how
a typical coherent phonon amplitude behaves as a function of laser
pulse width.  The formula indicates that the G band (LO mode)
intensity increases more rapidly than the RBM intensity, especially
when the laser pulse width is much shorter than each of the phonon
mode period.

\section*{Acknowledgments}
A.R.T.N acknowledges financial support from JSPS Fellowship for Young
Scientists.  R.S. acknowledges MEXT Grant No. 25286005.  The
University of Florida authors acknowledge NSF-DMR Grant No. 1105437
and OISE-0968405.  We are all grateful to Prof. J. Kono (Rice
University) and his co-workers for fruitful discussions which
stimulated this work.


\begin{thebibliography}{10}
\bibitem{note14-cp} It is apparent that the coefficient in front of
  the cosine term in Eq.~\eqref{eq:finalQ} is related to the Fourier
  transform of $u(t)$ given in Eq.~\eqref{eq:pulse}.  For a pulse
  shape in terms of Gaussian, the Fourier transform is also a
  Gaussian.
\end{thebibliography}

\end{document}